\documentclass[fleqn,10pt,lineno]{wlpeerj}
\usepackage{tabu}
\usepackage{diagbox}
\usepackage{verbatim}
\usepackage{url}

\title{Perspectives of Global and Hong Kong's Media on China's Belt and Road Initiative}

\author[1]{LE CONG KHOO}
\author[2]{ANWITAMAN DATTA}
\affil[1]{School of Computer Science and Engineering, Nanyang Technological University, Singapore}
\affil[2]{School of Computer Science and Engineering, Nanyang Technological University, Singapore}
\corrauthor[2]{ANWITAMAN DATTA}{ANWITAMAN@NTU.EDU.SG}

\begin{abstract}

This study delves into the media analysis of China's ambitious Belt and Road Initiative (BRI), which, in a polarized world, and furthermore, owing to the very polarizing nature of the initiative itself, has received both strong criticisms and conversely positive coverage in media from across the world. In that context, Hong Kong's dynamic media environment, with a particular focus on its drastically changing press freedom before and after the implementation of the National Security Law is of further interest. \\

Leveraging data science techniques, this study employs Global Database of Events, Language, and Tone (GDELT) to comprehensively collect and analyse (English) news articles on the BRI. Through sentiment analysis, we uncover patterns in media coverage over different periods from several countries across the globe, and delve further to investigate the the media situation in the Hong Kong region. This work thus provides valuable insights into how the Belt and Road Initiative has been portrayed in the media and its evolving reception on the global stage, with a specific emphasis on the unique media landscape of Hong Kong. \\

In an era characterised by increasing globalisation and inter-connectivity, but also competition for influence, animosity and trade-wars, understanding the perceptions and coverage of such significant international projects is crucial. This work stands as an interdisciplinary endeavour merging geopolitical science and data science to uncover the intricate dynamics of media coverage in general, and with an added emphasis on Hong Kong. 
\end{abstract}

\begin{document}
\nolinenumbers
\flushbottom
\maketitle
\thispagestyle{empty}

\section*{Introduction}
\subsection*{Motivation}
In this study, we foremost explore the evolving media coverage of the Belt and Road Initiative (BRI) on a global scale. The BRI, a global infrastructure development strategy launched by China in 2013 \citep{McBride_chinas_2023}, has garnered increasing media attention as it fosters economic connectivity and cooperation between China and other nations. The media plays a pivotal role in shaping public perceptions, making it vital to comprehend how their coverage of the BRI varies over geography and evolved over time \citep{mccombs_setting_2013}.

Besides the global landscape, given the drastic recent changes in Hong Kong, it stands out as a crucial area for examination. Specifically, the region has witnessed a noticeable decline in press freedom \citep{RSF_2023}, notably in the year 2022 \citep{Thomala_2023}, prompting us to delve deeper into Hong Kong news articles. The enactment of the National Security Law (NSL) \citep{NSL_2020} in 2020 further added complexity to the media landscape in Hong Kong, making it an essential lens through which to analyse the evolving narrative surrounding the BRI. 

Hong Kong, renowned as a global financial hub and a key gateway to China \citep{HKMA_2023}, plays a pivotal role in the success of the BRI. The initiative holds significant importance for Hong Kong, given its strategic geographical position and economic ties with China. Yet, despite the inherently political nature of the BRI, its sensitivity does not surpass the recent socio-political events experienced by Hong Kong, and in particular the BRI is inherently decoupled from the key domestic political issues in Hong Kong. This study thus allows us to uncover unique perspectives and implications of media discourse on BRI in general, but by furthermore focusing within a region experiencing dynamic socio-political changes, it also helps us understand some aspects of the nature and extent of changes in Hong Kong media. 

These thus constitute the two primary objectives and contributions of this work at a high level: understanding the geographic and temporal variety of media coverage of the Belt and Road Initiative (BRI), and drilling down on Hong Kong media's coverage of BRI to infer the impact of recent political changes on a topic which does not have immediate connect with the internal political issues, even as it aligns well with Hong Kong's role as a financial hub and thus its economic interests.

One valuable resource for analysing how media coverage of the BRI has changed and for determining how events have influenced these changes is the Global Database of Events, Language, and Tone (GDELT\footnote{https://www.gdeltproject.org/}). GDELT is a massive, open-source database that monitors and analyses worldwide news coverage, providing valuable insights into global events and their media coverage \citep{noauthor_gdelt_nodate}. An article "Can Big Data Stop Wars Before They Happen?" \citep{Himelfarb_2014} serves as a compelling testament to the significant impact of GDELT. It showcases how big data such as GDELT has emerged as a powerful resource in conflict prevention and mitigation, underscoring its potential to avert conflicts before they escalate. GDELT's sentiment analysis \citep{noauthor_introducing_nodate} is based on a machine learning algorithm that uses dictionaries containing lists of words and their corresponding sentiment values \citep{Saz-Carranza_2020}. This method allows for a more efficient calculation of sentiment scores for large datasets, as required by this study.

One study on the BRI \citep{garcia-herrero_countries_2019}, despite being confined to data from 2017-2018, presents a notable discovery. It unveiled that the BRI, while predominantly a politically focused initiative, is perceived less sensitively in terms of geopolitical implications. On the other hand, the second study \citep{dominic_2023}, which focuses on the European Union (EU) from 2017-2019, presents a comprehensive analysis of the trend within the EU region. However, to date, no study has delved into Hong Kong media's perspective on the BRI, especially in the recent years when Hong Kong has witnessed significant political transformations. This gap is critical given Hong Kong's pivotal role as a financial gateway into China and its importance regarding the BRI. Understanding how the BRI is portrayed in Hong Kong's media can provide a more balanced and nuanced perspective due to the city's unique position and the political changes it has experienced.

Therefore, part of this study centers on analysing Hong Kong's media coverage of the BRI. Utilising GDELT's capabilities, and augmenting it with further actual news content scraped directly from the news sites and employing sentiment and text analysis tools, we unveil prevalent sentiments and trends and infer whether and how internal political changes have impacted Hong Kong's media coverage of the BRI. This is augmented by our review of and comparison with the global scene comprising sentiment analysis of media coverage in several major BRI stakeholders, be it by the virtue of being a BRI participant, e.g., Pakistan and Malaysia, or by being prominent skeptics of BRI, e.g., United States, United Kingdom and India.


In the next section titled `Background Context', we provide short summaries of the Belt and Road Initiative and Hong Kong's deteriorating press freedom in the context of recent changes in its political circumstances. Next, in the `Data Curation' section we explain how we collect and clean the data used for this study. This follows a `Data Exploration' section where we report some basic statistics regarding our data set, and draw of initial inferences. We then delve into the application of `Sentiment Analysis'. Since most of our analysis is carried out with tonal score directly derived from our primary data source, GDELT, we commence with a benchmark of GDELT tonal scores against a more popular and mainstream sentiment analysis (TextBlob) to ensure that it is reasonable to use GDELT tonal scores. We augment the inferences from the sentiment analysis with further analysis of the actual textual content, investigating in particular certain specific aspects of Hong Kong's recent political issues and events. We then provide a saummary of key findings of our investigation and conclude.

\textbf{Data \& code:} The curated data, and the codes for data curation and analysis are available at \url{https://github.com/ChongLeh/gdelt-hongkong-media-analysis}.

\section*{Background context}

\subsection*{The Belt and Road Initiative}

\begin{figure}[ht!]
    \centering
    \includegraphics[width=0.95\textwidth]{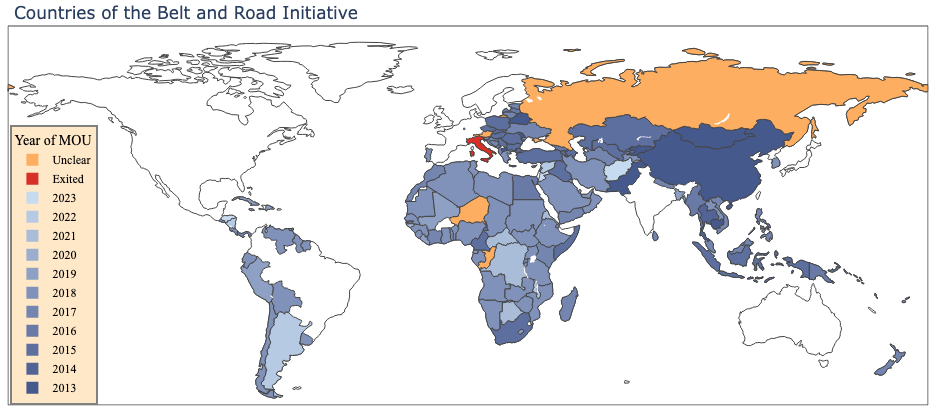}
    \caption[Partial Map of the BRI]{Partial Map of the BRI, recreated using data from \citep{Nedopil_2023}. Unshaded countries represent countries that have never/yet participated in the BRI.} 
    \label{fig:brimap}
\end{figure}

The BRI, also known as One Belt One Road (OBOR), is a global (see Figure \ref{fig:brimap}) infrastructure and economic development project launched by China in 2013. The initiative nominally aims to enhance connectivity, trade, and economic cooperation among countries across Asia, Europe, Africa, and beyond. It is comprised of two main components: the Silk Road Economic Belt, a land-based network of roads, railways, and pipelines, and the 21st Century Maritime Silk Road, a sea-based route linking key ports and maritime hubs \citep{EBRD}. 


\paragraph{Silk Road Economic Belt (SREB):}
The SREB seeks to revive and modernise historical trade routes, promoting economic cooperation, cultural exchange and development across multiple countries and continents. It primarily focuses on overland routes connecting China to Europe through Central Asia and Middle East \citep{Idikut2020}. It involves the construction of highways, railways, pipelines, and other critical infrastructure that facilitate the movement of goods, services, and people. As an integral part of the BRI, the SREB plays a vital role in shaping the geopolitical and economic landscape of the regions it traverses, complementing the efforts of the land corridors component. The land corridors connecting different regions include:
\begin{itemize}[noitemsep]
    \item \textbf{China-Mongolia-Russia Economic Corridor (CMREC)}: Connecting China, Mongolia, and Russia, this corridor enhances economic ties and trade between these northern Asian nations.
    \item \textbf{New Eurasian Land Bridge (NELB)}: Connecting China and Europe, traversing Central Asia and Russia, facilitating efficient rail and road transportation between the two continents.
    \item \textbf{China-Central Asia-West Asia Economic Corridor (CCWAEC)}: Connecting China, Central Asia and Middle East, facilitating economic and trade cooperation, focusing on the flow of capital to these regions.
    \item \textbf{China-Indochina Peninsula Economic Corridor (CICPEC)}: Connecting China with Indochina Peninsula, focused on the Association of Southeas Asian Nations (ASEAN) countries.
    \item \textbf{China-Pakistan Economic Corridor (CPEC)}: Connecting China and Pakistan, facilitating bilateral exchanges and cooperation between the two countries.
    \item \textbf{Bangladesh-China-India-Myanmar Economic Corridor (BCIMEC)}: Connecting China with South Asia, aimed at linking the two major markets of China and India.

\end{itemize}

\paragraph{21st Century Maritime Silk Road (MSR):}
The MSR focuses on maritime routes connecting China's coastal areas to Europe through the South China Sea, the Indian Ocean, and the Mediterranean Sea \citep{fmprc_2015}. This component aims to enhance maritime trade, port infrastructure, and maritime connectivity. It encompasses the development of ports, shipping lanes, and logistics hubs along key maritime routes. 

The BRI spans across regions that are of strategic importance and hold deep historical connections. Its impact on regions closely aligned with United States allies underscores the complex interplay between economic development, geopolitics, and diplomatic relationships. As China invests in infrastructure and connectivity in these regions, it opens avenues for collaboration, competition, and potential reconfigurations of global influence networks.

\subsection*{Hong Kong's Deteriorating Press Freedom}

\begin{figure}[htbp!]
    \centering
    \includegraphics[width=0.95\textwidth]{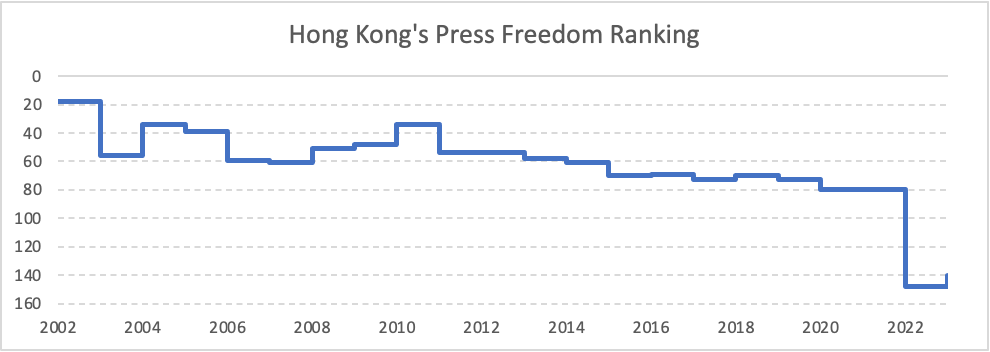}
    \caption[Hong Kong Press Freedom Ranking Graph]{Hong Kong Press Freedom Ranking Graph, reproduced based on \citep{Reuters}}
    \label{fig:hkgraph}
\end{figure}

Historically, Hong Kong witnessed a relatively free and vibrant media landscape \citep{Lee_1998}, with notable levels of press freedom even after the 1997 handover from British rule to Chinese sovereignty. Press freedom remained relatively high, as indicated by its rank of 18 in the 2002 index \citep{RSF_2002}. However, in recent years, particularly in the backdrop of shifting political dynamics and increased influence from mainland China, there has been a noticeable decline in press freedom (see Figure \ref{fig:hkgraph}). This has been effectuated both by subtle tactics such as taking over of the management of media companies by companies with close links to the Chinese government purchasing majority shares \citep{Lhatoo_2016}\citep{The_Standard_2021}, and also by more explicit means by the implementation of various laws such as the national security law in Hong Kong, enacted in 2020 (more on these below). All in all, the influence of China has notably increased, impacting the editorial independence and overall tone of media outlets in Hong Kong. In response to the evolving political climate and increased influence from China, reporters in Hong Kong have increasingly resorted to self-censorship \citep{RSF_2023a}. Fearing repercussions, they may refrain from covering sensitive topics or expressing critical views on the government or Beijing's policies. This self-imposed limitation on reporting has altered the previously diverse and open media landscape, affecting the breadth and depth of information available to the public.

\subsubsection*{Legislation \& Regulations}

\paragraph{National Security Law (2020):}
Enacted on June 30, 2020, by the Standing Committee of the National People's Congress of China \citep{NSL_2020}, this law criminalises secession, subversion, terrorism, and collusion with foreign forces. Its implementation has had a profound impact on media organisations, directly targeting reporters \citep{RSF_2023b} and news outlets \citep{BBC_2020}\citep{Grundy_2020}, severely limiting their ability to freely report on certain topics.

\paragraph{Use of Sedition Law (2020):}
Sedition laws have a historical context in Hong Kong dating back to its colonial era under British rule. The Crimes Ordinance was initially enacted during the British colonial period \citep{Wong_2008}, primarily to suppress any acts perceived as subversive or threatening to the colonial authorities. This law was retained and carried forward after Hong Kong's transfer of sovereignty to China in 1997. However, the significant shift occurred with the implementation of the National Security Law in 2020, imposed by Beijing. Since the handover, the first person to be charged under the sedition law was a Hong Kong opposition activist, Tam Tak-chi \citep{Lau_2020}. Subsequently, the utilisation of the sedition law gained momentum \citep{Law_2023}, often paralleling the application of the more comprehensive National Security Law. Instances of its use included the indictment of a 23-year-old Hong Kong student who published "seditious" comments online during her stay in Japan, demonstrating how the law is applied even to people residing overseas, thus  potentially limiting press freedom \citep{Ye_2023}. 

\paragraph{Electoral Reforms (2021):}
The Electoral Reforms of 2021 \citep{zh_2021} significantly impacted Hong Kong's political landscape and had consequential implications for press freedom. With directives from the central government in Beijing, the reforms reduced directly elected seats in the Legislative Council (LegCo) and amplified the influence of pro-Beijing committees like the Election Committee \citep{Tian_Zaharia_2021}. The introduction of a "patriotism requirement" for candidates emphasised Beijing's increasing influence \citep{bloomberg_2021}, signaling a tighter grip on the political narrative. These changes stirred concerns about the erosion of democratic principles and freedoms, amplifying the ongoing debate about Hong Kong's autonomy and its integration within the broader political framework of China, which includes the realm of press freedom.


\section*{Data curation}

\subsection*{Candidate Data Sources}

We first describe the considerations based on which we finally chose GDELT \citep{noauthor_gdelt_nodate} as the primary source of data to drive this study. 

\paragraph{Google News API:}
The Google News API is a data interface provided by Google that grants access to a collection of news articles from diverse sources worldwide, facilitating real-time updates and enabling researchers to retrieve and analyse media coverage on various topics. However, it has certain limitations that should be considered when conducting media analysis on the BRI. First, the API's news sources are limited to media outlets accepted and indexed by Google, potentially excluding niche or regional news sources. This restriction may result in an incomplete view of media coverage on the topic.

Moreover, the availability of news articles on the API could be influenced by language and region limitations. Despite covering a wide range of languages and regions, there might be gaps in coverage, particularly for less commonly spoken languages or specific geographic areas related to the BRI.

Another notable limitation is the API's reliance on Google's algorithms to select and rank news articles \citep{GoogleNews_rank}. These algorithms determine the relevance and prominence of news stories, which could introduce inherent bias and limit control over article inclusion in search results. This aspect may impact the comprehensiveness and neutrality of the data retrieved.

Additionally, the Google News API typically provides only snippets of news articles, lacking access to the full-text content. This constraint might hinder in-depth analysis that requires a complete understanding of context and nuances present in the articles. 

Nevertheless, the Google News API offers significant advantages, such as real-time updates, enabling access to the latest news articles related to the Belt and Road Project. Furthermore, it is an easy-to-use data source, with the convenience of accessing and integrating data through Python APIs \citep{Hu_2019}.

\paragraph{News-please:}
News-Please \citep{Hamborg2017} offers researchers three distinct modes: CLI mode, library mode, and news archive mode from CommonCrawl.org. Each mode provides unique advantages and challenges for media analysis.

In CLI mode, researchers can leverage the Command-Line Interface to perform comprehensive web scraping, extracting news articles from various websites. This mode requires manual adding of root URLs of news outlets or extensive configuration to function properly, which is challenging for large-scale datasets like media articles on the BRI. This is also a challenge when using the library mode.

In news archive mode, it uses CommonCrawl's web archive to crawl and extract data. However, due to the sheer size and scale of the web archive, crawling and extracting relevant news article from this massive dataset can be both expensive and time-consuming.

In conclusion, while News-Please offers a potentially rich and extensive dataset for media analysis through its various modes, it requires either time-consuming manual work or crawling and processing of extremely large dataset.

\paragraph{GDELT:}
GDELT, the Global Database of Events, Language, and Tone, provides an extensive collection of news articles for media analysis, from 1979 to the present \citep{noauthor_gdelt_nodate}. GDELT's database consist of articles from more than 100 different languages. The database can be used through the GDELT Analysis Service, Google BigQuery, or used directly using the raw data files.

The GDELT Analysis Service is cloud-based and is used through a variety of tools and services offered by GDELT to visualise or explore the database. These tools have a fixed set of functions based on their Event Database (EVENT) or Global Knowledge Graph (GKG). It offers a collection of structured events extracted from their database, including attributes such as actors, location, time and event type \citep{GDELT_AS}. However, as it is a centralised set of tools, the functions are limited and the BRI is not part of the keywords available in the GDELT Event Database. Similarly, the Google BigQuery is also restricted by the Event keywords, although this method allows queries dating back to 1979. Lastly, for the raw data files, they can be accessed through multiple ways, including the GDELT DOC 2.0 API. 

\paragraph{Data source used:} For this work, we ultimately chose the GDELT DOC 2.0 API \citep{GDELT_DOCAPI} due to its user-friendly nature and flexibility. This approach allows us to utilise a substantial volume of news articles while following a straightforward and efficient workflow. Although the API's search timeframe is limited to 2017 onwards, the period from 2017 to the present offers a sufficient span to identify any pertinent trends within the current political landscape. Despite the BRI's commencement in 2014, this duration is sufficient to enable a comprehensive analysis that aligns with the prevailing dynamics in the political sphere.

Additionally, the GDELT DOC 2.0 API offers its own built-in sentiment analysis tone scores, providing an immediate and valuable resource for initial insights. Despite the basic nature of the underlying sentiment analysis method \citep{Saz-Carranza_2020}, these scores offer a starting point for the research. The use of these scores also reduces the chances of missing articles, especially older articles, which can be difficult or impossible to scrape and analyse. Moreover, these scores can be utilised for comparative analysis alongside other sentiment analysis tool (such validation using statistical tests is reported subsequently in the section on `Sentiment Analysis'). The GDELT data was complemented with some data we scraped directly from online news sites (described below), which was then leveraged for both data cleaning and sentiment analysis validation. 

\subsection*{Data Collection}

\begin{figure}[ht!]
    \centering
    \includegraphics[width=0.95\textwidth]{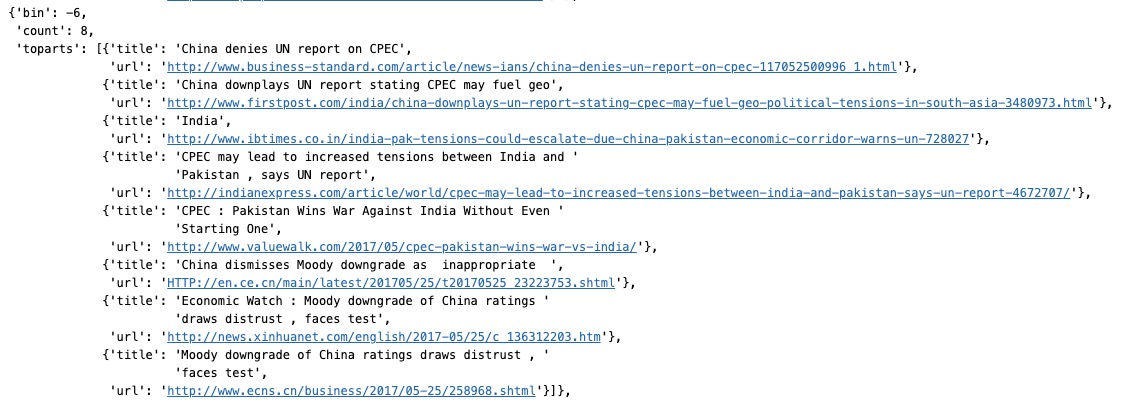}
    \caption{Example API request output: Articles with a tonal score of -6 (indicated with `bin').}
    \label{fig:apioutput}
\end{figure}

The GDELT DOC API 2.0 allows the collection of data in different modes and query types. We used an option called `ToneChart', and looked for all `English' language articles containing the search terms `belt and road' and `one belt one road', which respectively represent the older and newer denominations of the BRI, iterated through all the days from 2017 till 07/09/2023. Besides the title and article URL for the supposed relevant articles for a given date (which is part of the query), owing to the choice of `ToneChart' in the query, a tonal score (as precomputed by GDELT's proprietary sentiment analysis algorithm) for the articles is also obtained in (a JSON encoded) response to the query, which groups the response by this tonal score. As an example, in Figure \ref{fig:apioutput} we show the response obtained for 25/05/2017 corresponding to the tonal score (indicated with `bin') of -6. 

\paragraph{Other languages:}
Since GDELT encompasses news articles in languages beyond English, we aimed to retrieve these articles too. To evaluate the added value of this supplementary dataset, we specifically examined NATO countries as a practical case study, with the number of relevant articles per country per year summarized in Table \ref{tab:natooutput} (in the Appendix).

We noted that there were significant news articles for certain countries in 2017. However, this count gradually diminishes to zero by 2023. This could potentially indicate a waning interest in the BRI over time. The lack of news articles for other countries in other languages may also be attributed to our limitations in using GDELT's search to accurately capture the precise keywords, since our own queries were still confined to the English phrases such as "One Belt One Road" or "Belt and Road," leading to a potential under-representation of related articles. Given these concerns regarding representativeness or sparsity of information in non-English languages, we have chosen to confine this study exclusively to English news articles. This decision is also influenced by the potential complexity that incorporating various languages could introduce during the later stages of further analysis.

\subsection*{Data collection for subsequent tonal score validation and more detailed analysis of articles in Hong Kong media}

Since GDELT's proprietary sentiment analysis tool is not mainstream and the details of the underlying algorithm are not well documented, we carried out a micro benchmark. To that end, we used the \textit{newspaper3k} library \citep{Ou-Yang_2018} to scrape the actual article contents from the URLs obtained through GDELT (as discussed below, this was inadequate in certain instances, in which case we attempted to use `BeautifulSoup' \citep{richardson2007beautiful} and `requests' \citep{reitz2011requests}). Both because of our objective to drill down in Hong Kong, and to keep the tonal score benchmarking process tractable, we confined this to Hong Kong specific articles (we discuss later how we determine the geographic label for articles/media domains). The initial web scraping process resulted in a distribution of articles summarized in Table \ref{tab:scrapearticles}.
 \begin{table}[!ht]
    \centering
    \footnotesize
    \begin{tabu}{|l|[2pt]l|l|l|l|l|l|l|l|}
    \hline
        Year    & 2017 & 2018 & 2019 & 2020 & 2021 & 2022 & 2023 & \textbf{Total} \\ \tabucline[2pt]{-}
        Invalid & 116  & 84   & 21   & 6    & 12   & 16   & 15  & \textbf{270}  \\ \hline
        Valid   & 574  & 602  & 216  & 49   & 60   & 48   & 67  & \textbf{1616}  \\ \hline
        Total   & 690  & 686  & 237  & 55   & 72   & 64   & 82  & \textbf{1886} \\ \hline
    \end{tabu}
    \caption{Distribution of Hong Kong articles scraped using \textit{newspaper3k}}
    \label{tab:scrapearticles}
\end{table}

Articles with less than 40 words in their text bodies were categorized as invalid (we designate these as `non-article'). This criterion served as a standard to identify and filter out articles lacking substantial content for further analysis, ensuring efficiency and accuracy in the subsequent automated sentiment analysis benchmarking. Additionally, this statistics facilitated identifying potential reasons for scraping failures, allowing for targeted efforts to recover these articles through additional scraping or analysis.

Furthermore, manually examining the overall corpus of invalid articles from the initial scraping provided valuable insights. The process resulted in the following distribution (summarized in Table \ref{tab:invalidart}) of invalid articles and the contributing factors.

\begin{table}[!ht]
    \centering
    \footnotesize
    \begin{tabu}{|l|l|l|}
    \hline
        Domain & Count & Factors\\\tabucline[2pt]{-} 
        7thspace.com            & 158   & Missing Article(s) \\ \hline
        ejinsight.com           & 50    & Missing Article(s) \\ \hline
        hongkongherald.com      & 43    & Scrape Error       \\ \hline
        scmp.com                & 14    & Non-article        \\ \hline
        rthk.hk                 & 2     & Non-article        \\ \hline
        chinaeconomicreview.com & 1     & Missing Article(s) \\ \hline
        prestigeonline.com      & 1     & Missing Article(s) \\ \hline
        squarefoot.com.hk       & 1     & Domain moved       \\ \hline
    \end{tabu}
    \caption{Distribution of invalid articles by domain and factors}
    \label{tab:invalidart}
\end{table}

For Hong Kong Herald articles, we could thus 
use direct \textit{requests}' and '\textit{BeautifulSoup}' to retrieve 41 of the relevant articles, however, two articles could still not be scraped as they were missing from the source. The additional web scraping process thus resulted in the final distribution of articles as summarized in Table \ref{tab:finalscrapeearticles}.
 \begin{table}[!ht]
    \centering
    \footnotesize
    \begin{tabu}{|l|[2pt]l|l|l|l|l|l|l|l|}
    \hline
        Year    & 2017 & 2018 & 2019 & 2020 & 2021 & 2022 & 2023 & \textbf{Total} \\ \tabucline[2pt]{-}
        Invalid & 116  & 82   & 19   & 1    & 4    & 1    & 6   & \textbf{229}  \\ \hline
        Valid   & 574  & 604  & 218  & 54   & 68   & 63   & 76  & \textbf{1657} \\ \hline
        Total   & 690  & 686  & 237  & 55   & 72   & 64   & 82  & \textbf{1886} \\ \hline
    \end{tabu}
    \caption{Final distribution of scraped Hong Kong articles}
    \label{tab:finalscrapeearticles}
\end{table}


\subsection*{Data Cleaning}

\paragraph{Data Labelling:}
As evident from Figure \ref{fig:apioutput}, each article is accompanied by its corresponding tonal score, title, and URL. Consequently, each news article still needs to be labelled with their respective source countries. 

For each unique media domain in the overall corpus of news articles previously obtained, utilising the GDELT DOC 2.0 API once more, but with `ArtList' mode, for a vast majority of the articles we could automatically determine the country associated with the media domain (provided the GDELT database had this information). As evident from the summary presented in Table \ref{tab:domainlabelresults}, the application of the API successfully labelled over 165,447 news articles. However, a total of 2,270 domains had remained unlabelled, and undertaking manual labelling for all these would have been prohibitive. To address this challenge, we divided the unlabeled articles across two groups, those from domains with at least 50 articles, or less. For the former, we manually determined the country of the domain, thus resolving the label for more than half of the entries for which the automated process was inadequate. This approach substantially diminished the need for manual data labelling, thereby minimising associated costs, striking a pragmatic balance. 
\begin{table}[ht]
    \centering
    \footnotesize
    \begin{tabu}{|l|[2pt]l|l|l|}
        \hline
        \diagbox[width=\dimexpr \textwidth/4+4\tabcolsep\relax, height=1cm]{ Domains }{Category}
                           & Number of Articles & Number of Domains & Labeling status\\ \tabucline[2pt]{-}
            Labelled by API & 165447 & 4155 & Automatic\\ \hline
            50 or More Articles & 12689 & 62 & Manual\\ \hline
            Fewer than 50 Articles & 8967 & 2208 & Unlabeled\\ \hline
    \end{tabu}
    \caption{Summary of Domain Labelling Results}
    \label{tab:domainlabelresults}
\end{table}


\paragraph{Repetitive articles:}
The originally scraped dataset included repetitive articles, particularly from China's media outlets, where various domains repost the same news articles. This repetition can distort the tonal score for a country, and of the average computed over the global corpus due to the duplicated content. To address this issue, articles with (almost) identical titles were eliminated. This process involved removing news articles with the same first 5 words in the title, as the latter part often contains the name of the respective media domain. While this method would not identify all repeated articles, it offers a reasonable compromise compared to manual verification. As a result of this process, a total of 94,781 news articles were removed. This resulted in 102,552 news articles to study.

\paragraph{Invalid media domains:}
To enhance dataset cleanliness, we applied additional refinement steps eliminating some questionable media domains using the following intuitive ad-hoc criteria:

\begin{itemize}
    \item Repost-Focused Domains: We identified and discarding domains that predominantly repost news articles (as identified above) from other sources, potentially skewing data duplication.
    \item Non-BRI Content: We removed domains with a consistent record of publishing articles unrelated to the BRI that were still flagged as such by GDELT's algorithm. The identification process involved manual review of the URLs, revealing a prevalence of low-quality websites where GDELT indiscriminately included every article from these domains.
    \item Non-Media Blogs: We excluded Wordpress and other easy to identify blogging platforms that lack characteristics of established media outlets.
\end{itemize}

\paragraph{Further data cleaning:}
Among the records for 2017, 'news.xinhuanet.com' represented a substantial one-third of the articles. Based on web scraping for further analysis, we identified many non-BRI 'news.xinhuanet.com' content being mislabeled by GDELT, an issue  encountered previously for a few other low-quality data sources. This approach successfully facilitated the removal of a significant majority of 'news.xinhuanet.com' articles, specifically 4952 out of 5061 were determined irrelevant.

As a results of these processes, combining the elimination of repetitive articles and domains with insignificant number of news articles, the final resulting dataset which we use to drive the study reported in the rest of this paper comprises 85,982 news articles from 3,485 domains across 161 countries. 

\section*{Data Exploration}
\subsection*{News Article Counts}

\begin{figure}[ht!]
    \centering
    \includegraphics[width=0.95\textwidth]{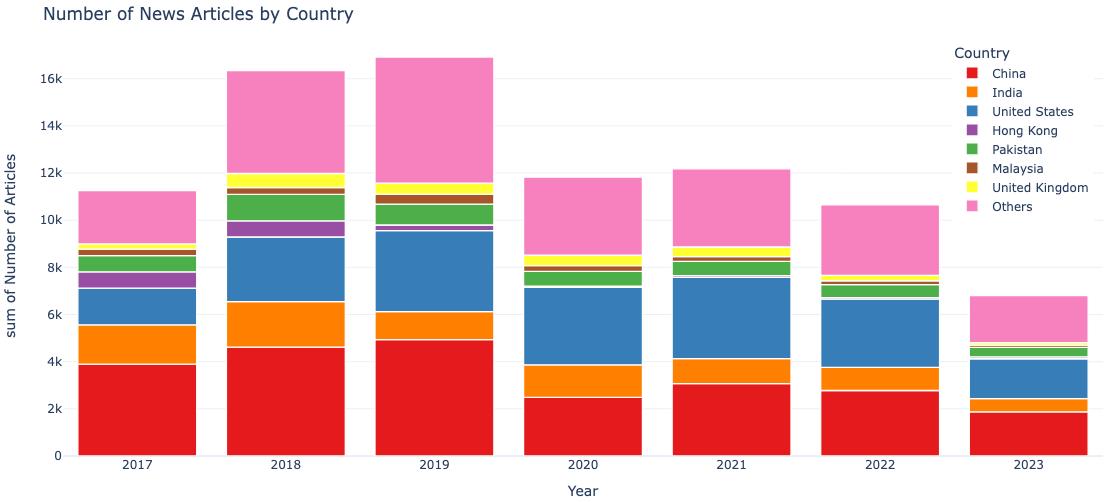}
    \caption{Bar chart of news article counts by country}
    \label{fig:articlecount}
\end{figure}

Figure \ref{fig:articlecount} shows the distribution of the number of articles from various countries over the years. China and the United States hold the highest count of news articles across the years, displaying a dominant presence. India, Pakistan, United Kingdom, and Malaysia follows, maintaining a substantial but lesser count compared to China. Hong Kong, while not at the forefront in terms of article count, remains a significant contributor, especially given its status both as a global financial hub and its role as a special administrative region in China. These countries/regions\footnote{Disclaimer: We use the term `country' loosely in this paper, for brevity. In the context of Hong Kong, we note that it is in fact a special administrative region in China.} contribute the majority of the yearly article count, showcasing their influence and active participation in news coverage, particularly in the context of the BRI.\\

\subsection*{Average Tone}

\begin{figure}[h!]
    \centering
    \includegraphics[width=0.95\textwidth]{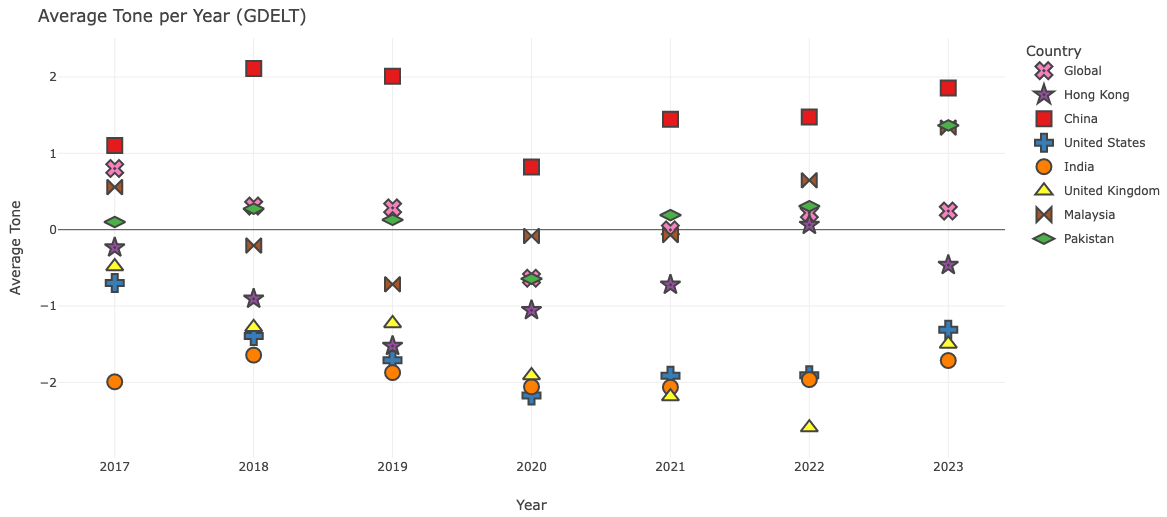}
    \caption{Average tone (GDELT) per year by country}
    \label{fig:tonegdelt}
\end{figure}

Using GDELT's sentiment score, a prevalent pattern emerged across the years leading up to 2020 (see Figure \ref{fig:tonegdelt}). For most countries, including the global trend\footnote{Disambiguation: Given the strong polarization on the topic, there is not always an actual universal `global trend', but we use this term to refer to the cumulative score computed from the whole corpus of articles considered.}, there was a noticeable downward trajectory in average tone. This decline indicates a shift towards more negative or critical reporting in the media.

However, Hong Kong presented a different narrative. Until 2019, it followed the general declining trend, aligning with the prevalent pattern observed in other countries. However, in 2020, Hong Kong took a distinctive turn, showcasing a more positive tone compared to the previous year. This marked deviation from the global and other national trends (including even within China) highlights a unique media response in Hong Kong during a pivotal year.

Since 2021-2022, most countries saw a slight increase in their average tone (the United Kingdom remained an exception, maintaining a downward trend till 2022). Nevertheless, India has consistently maintained one of the most negative tones on the BRI, while we see that other big skeptics, US and UK converge with India over time, and is significantly more negative than they were in 2017. In contrast, while Hong Kong was close to UK and US in 2017 and to certain extent, even in 2018, Hong Kong has largely diverged from UK and US since. Yet, this divergence is more a consequence of increased polarization among the pro/anti-BRI camps, while the average tone for Hong Kong, after an intermediate dip, is closer to where it was originally around 2017, demonstrating that the Hong Kong media continues to have a more neutral tone overall when it comes to the coverage of the BRI.

\subsection*{Hong Kong Media Focus: Most Frequent Words}

\begin{figure}[h!]
    \centering
    \includegraphics[width=0.95\textwidth]{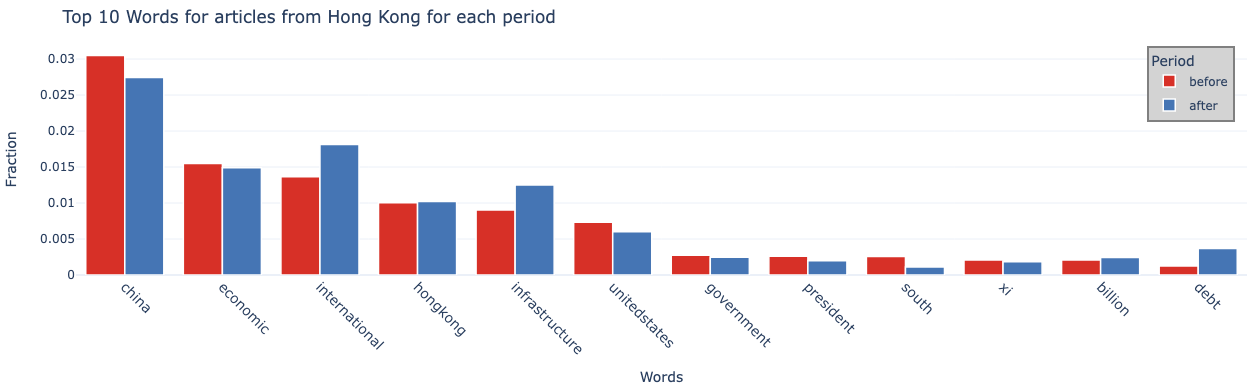}
    \caption{Top 10 words from Hong Kong News Articles (before: 2017-2019; after: 2020-2023)}
    \label{fig:hkword}
\end{figure}

Figure \ref{fig:hkword} illustrates the evolution of media focus in Hong Kong over the years. Since the number of articles are very different across the two periods: 1392 for 2017-2019 (before) vs 224 for 2020-2023 (after), the relative (ordering) of the word counts expose some interesting additional insights (the figure shows a union of some of the topmost words from each of the two periods). While the overall focus remains consistent, there are notable shifts worth mentioning. The most striking observation is the increased mention of 'debt'.  

\subsection*{Distribution of News Articles}

This section focuses on the distribution of sources of news articles across the top countries by news articles count, highlighting the primary news domains year by year offering insights into the prominence of specific domains within the different regions. 

\begin{figure}[h!]
    \centering
    \includegraphics[width=0.95\textwidth]{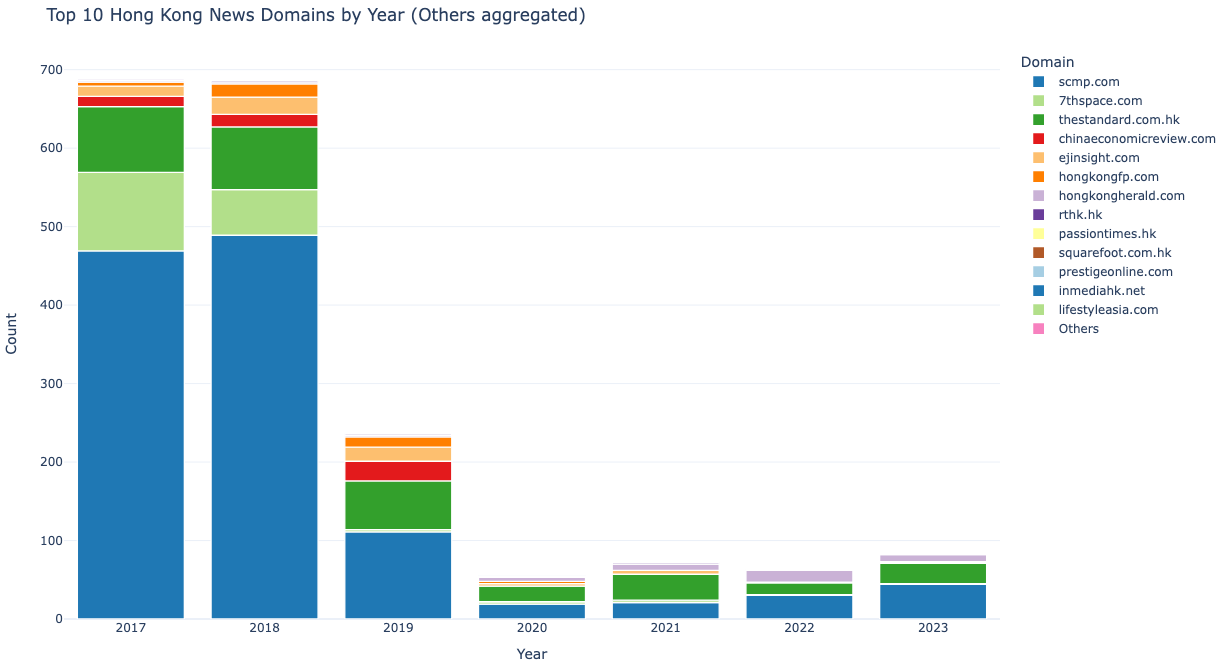}
    \caption{Articles Distribution for Hong Kong}
    \label{fig:hkdist}
\end{figure}

\paragraph{Hong Kong:}
The years 2017 and 2018 emerge as significant peaks in coverage (Figure \ref{fig:hkdist}), followed by a notable decline in 2019. The South China Morning Post (SCMP) takes the lead in covering the BRI during the peak years, closely followed by 7thspace and The Standard. However, despite the relative dominance of SCMP and The Standard in subsequent years until 2023, it is notable that the actual volume of coverage is heavily attenuated and the diversity of unique domains reporting on the BRI remains significantly low in Hong Kong compared to other countries. Given the multilingual nature of the region, incorporating diverse linguistic mediums beyond English might offer a more comprehensive understanding of Hong Kong's involvement and perspectives regarding the BRI, nevertheless, it is apparent that BRI as a topic has by and large become marginal in the recent years in Hong Kong media. 

\begin{figure}[h!]
    \centering
    \includegraphics[width=0.95\textwidth]{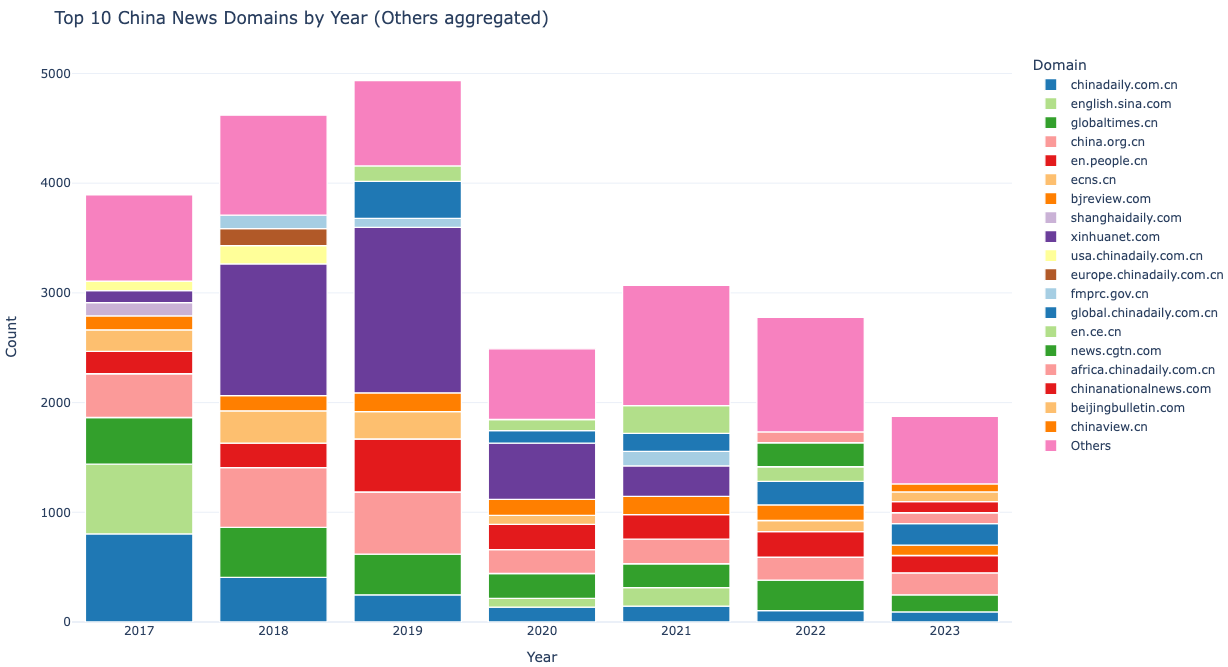}
    \caption{Articles Distribution for China}
    \label{fig:cndist}
\end{figure}

\paragraph{China:}
From 2018 to 2020 XinHua consistently dominates the coverage on the BRI (Figure \ref{fig:cndist}). Notably, the presence of unique top domains covering the BRI each year remains relatively low, with minority representation from domains outside the top 10. This observation suggests a limited diversity of voices and coverage within the Chinese media landscape concerning the BRI. Such a limited number of sources is not unexpected, considering both the lack of press freedom in China\footnote{As per Reporters Without Borders' 2023 press freedom index, China is ranked 179 among 180 countries (https://rsf.org/en/country/china).} and that we only use English articles. 

\begin{figure}[h!]
    \centering
    \includegraphics[width=0.95\textwidth]{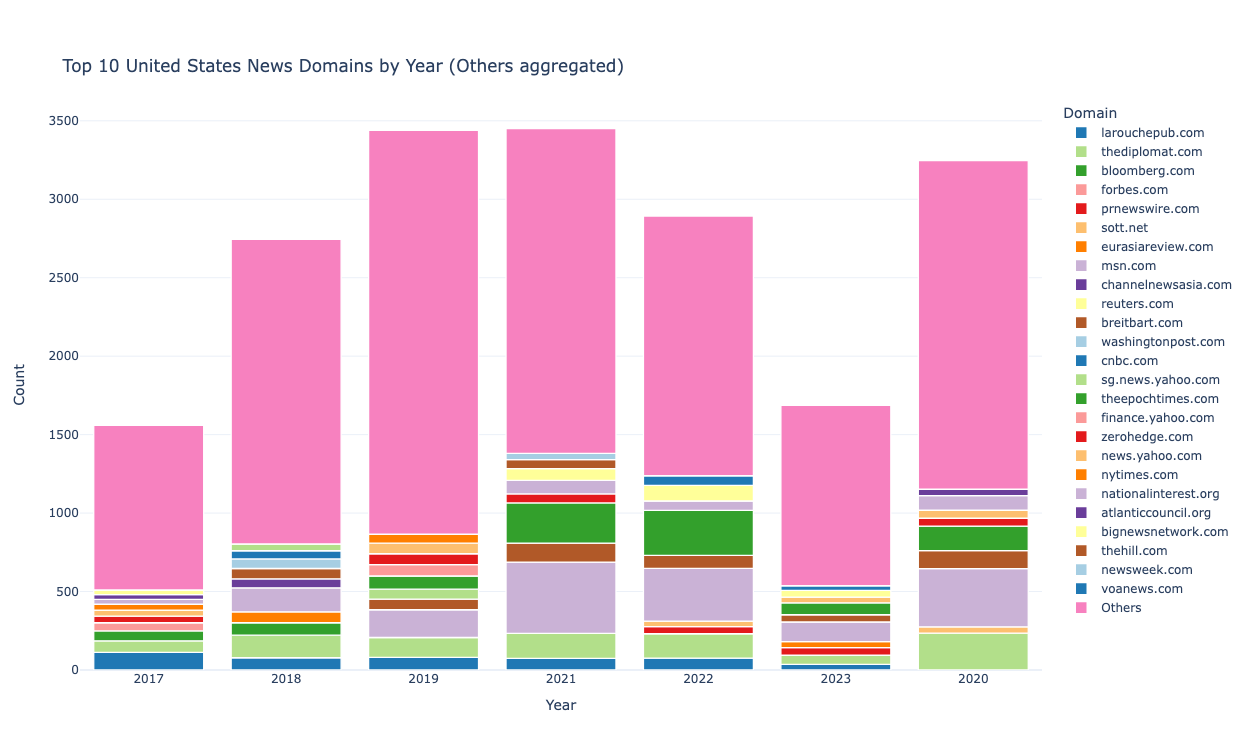}
    \caption{Articles Distribution for United States}
    \label{fig:usdist}
\end{figure}

\paragraph{United States:}
A noteworthy observation regarding the United States (Figure \ref{fig:usdist}) is the diverse range of unique top domains covering the Belt and Road Initiative (BRI) across various years. This diversity indicates a fluctuation in dominant coverage by different domains throughout different periods. Additionally, the considerable presence of other domains outside the top 10 highlights a notable diversity in media coverage, illustrating the breadth of perspectives across various media outlets. This trend aligns with expectations, considering the extensive number of media companies within the US and the presence of a free media environment. It underscores the dynamic and multifaceted nature of media coverage in the US regarding this global initiative, reflecting the multitude of viewpoints and narratives present within the American media landscape.
\clearpage
\begin{figure}[!h]
    \centering
    \includegraphics[width=0.95\textwidth]{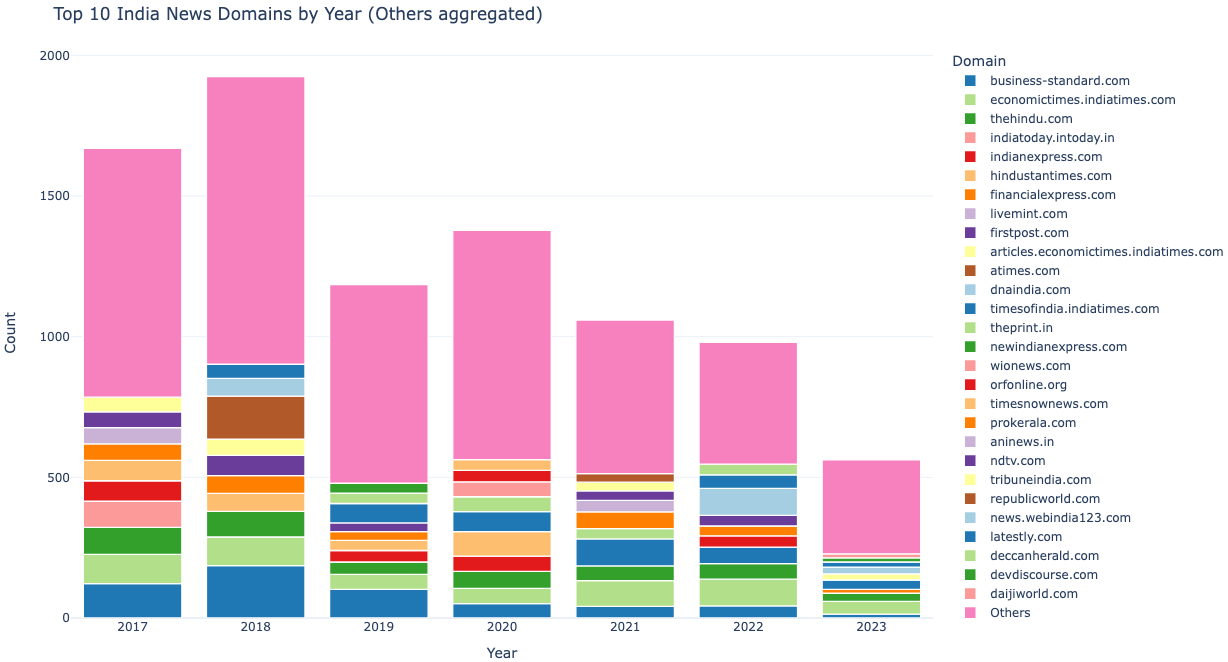}
    \caption{Articles Distribution for India}
    \label{fig:indist}
\end{figure}

\paragraph{India:}
India's scenario (Figure \ref{fig:indist}) mirrors that of the United States, with a significant portion of other domains outside the top 10, signifying a diverse array of sources contributing to the BRI coverage. Moreover, India showcases a similar trend of featuring numerous distinct top 10 domains, indicating a varied landscape of media outlets participating in the coverage of this global initiative.

\begin{figure}[h!]
    \centering
    \includegraphics[width=0.95\textwidth]{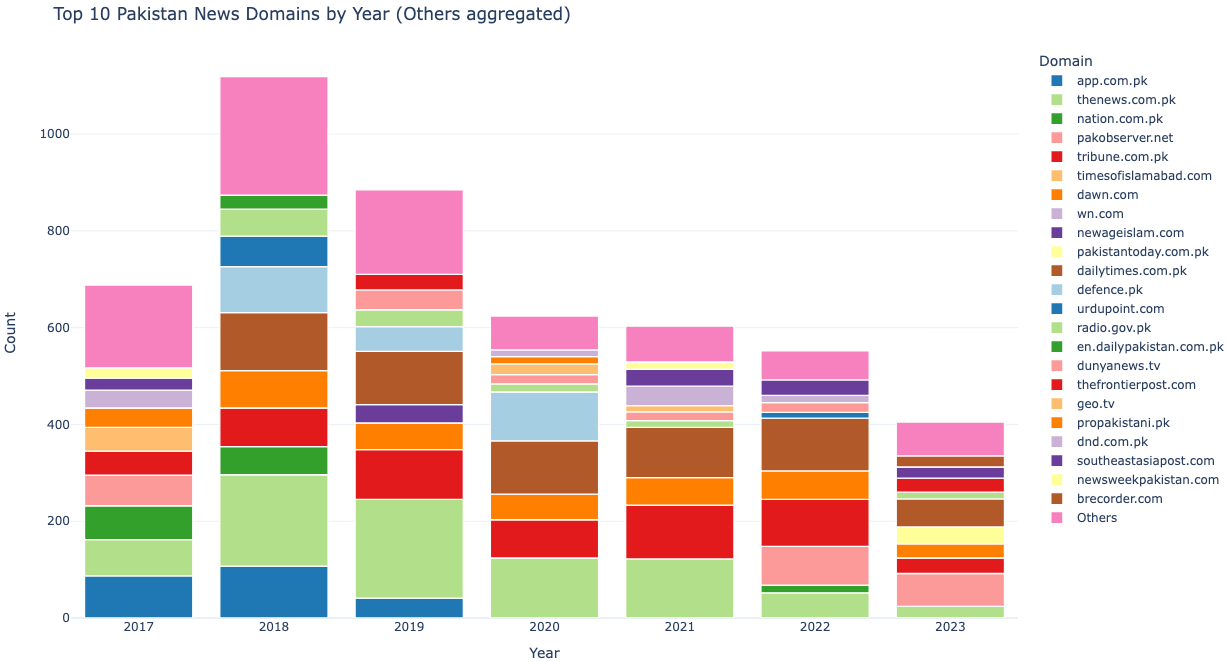}
    \caption{Articles Distribution for Pakistan}
    \label{fig:pkdist}
\end{figure}

\paragraph{Pakistan:}
In Pakistan's case (Figure \ref{fig:pkdist}), although the `others' category constitutes a smaller portion, the presence of diverse top 10 domains indicates a notable variety in the leading sources covering the BRI. The limited representation in the `others' category might reflect Pakistan's media landscape, which is less developed due to various factors intrinsic to the country's nature. This suggests that while there is less overall representation from various sources, the top domains in Pakistan offer diverse coverage of the BRI within the country's limited yet vibrant media landscape.

\begin{figure}[h!]
    \centering
    \includegraphics[width=0.95\textwidth]{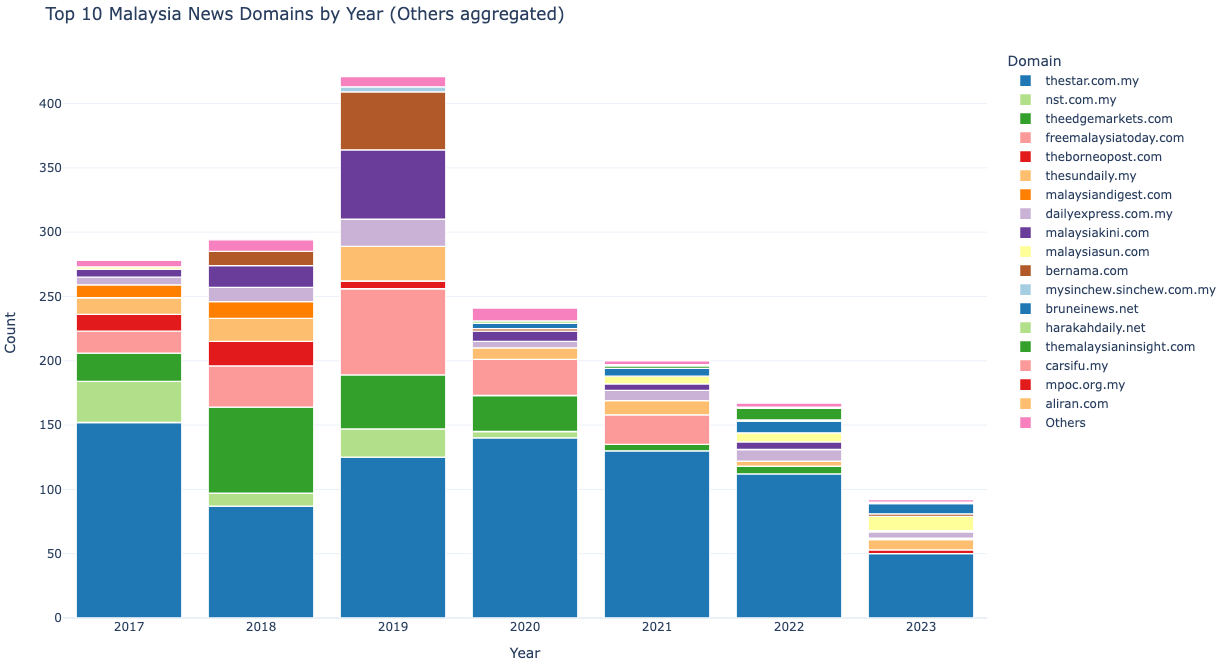}
    \caption{Articles Distribution for Malaysia}
    \label{fig:mydist}
\end{figure}

\paragraph{Malaysia:}
In Malaysia's case (Figure \ref{fig:mydist}), the `others' portion comprises almost a negligible fraction, indicating a minimal presence of sources beyond the top domains covering the BRI. Throughout all the years analysed, The Star consistently emerges as the primary contributor. As highlighted previously, the dataset's focus on English-language news articles may explain The Star's dominance, given its status as one of the most widely read English news domains in Malaysia, potentially influencing BRI coverage representation within Malaysia. 

\begin{figure}[h!]
    \centering
    \includegraphics[width=0.95\textwidth]{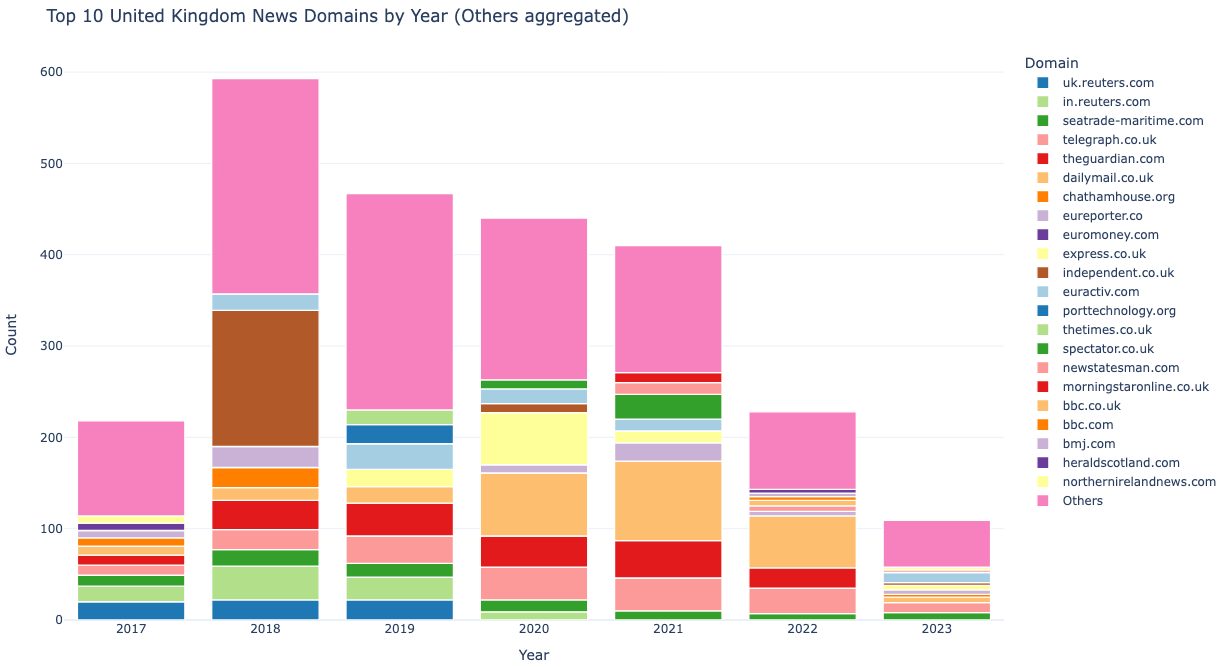}
    \caption{Articles Distribution for United Kingdom}
    \label{fig:ukdist}
\end{figure}

\paragraph{United Kingdom:}
In the United Kingdom's context (Figure \ref{fig:ukdist}), a similarity with the United States and India emerges in the diverse representation of top 10 domains covering the Belt and Road Initiative (BRI). Additionally, a substantial presence of `others' indicates a significant diversity beyond these top domains.

Overall, Countries where English serves as the primary language, like the USA and UK as well as countries where a large section of the society uses English as a lingua franca, e.g, India and Pakistan, exhibit diverse coverage of the BRI in their English language media. Among the countries which fall on the lower end in the Press Freedom Index, India stands out, displaying a distinct level of diversity in the media sources covering BRI. This deviation could potentially be attributed to India's substantial population size, and political and geographic diversity contributing to a more varied media landscape notwithstanding the deteriorating position in the press freedom index and subsequently, a more diverse range of perspectives on the BRI, which mostly does not have an immediate connection to India's internal domestic political issues.

\section*{Sentiment analysis}

\subsection*{Validation of GDELT tonal score}
GDELT offers valuable insights into media tone through its provided score, allowing a quantitative assessment of news sentiment. However, the sentiment analysis used by GDELT is a rather simplistic approach, quantifying tone primarily based on the balance between positive and negative words within the text \citep{Saz-Carranza_2020}, this method is also not documented in detail. 

To augment our credibility and provide a more comprehensive analysis of sentiment, we thus complement GDELT's tone assessment with the TextBlob sentiment analysis tool \citep{Shah_2020} which is a widely adopted Python library. TextBlob assigns scores to word collections based on a predefined dictionary of positive and negative words besides incorporating pretrained models, and is considered better suited for the analysis of formal language as expected in news articles, in contrast to (another option we considered but eventually did not use, namely) Valence Aware Dictionary and Sentiment Reasoner (VADER), which employs a lexicon-based approach, and provides polarity scores by considering context, emojis, punctuation \citep{Hutto_Gilbert_2014} and is known for its good performance in analysing sentiment across social media content \citep{Tymann2019GerVADERA}. This approach adds an additional layer of credibility to our study, especially considering that the majority of our experimentation relies on GDELT's tone sentiment.


\subsection*{Sentiment Analysis Correlation for Hong Kong News Articles}
\paragraph{Research Area:}
We seek to explore the degree of correlation between the average tone scores derived from GDELT and TextBlob sentiment analysis using the actual Hong Kong news articles that we had additionally scraped directly based on the URLs obtained from GDELT.

\paragraph{Null Hypothesis (H0):}
There is no significant correlation between the average tone scores obtained from GDELT and TextBlob sentiment analysis for Hong Kong news articles.

\paragraph{Alternative Hypothesis (H1):}
There is a significant positive correlation between the average tone scores obtained from GDELT and TextBlob sentiment analysis for Hong Kong news articles.

\paragraph{Methodology:}
Prior to assessing the correlation between the average tone scores derived from GDELT and TextBlob sentiment analysis in the context of Hong Kong news articles, it is imperative to conduct some analysis. The first step involves verifying the normality of the datasets. This normality check is crucial to determine whether the data conforms to a Gaussian distribution. The Shapiro-Wilk test \citep{shapiro_wilk_1965} will be used to test for normality. If both datasets passes the normality test and the Levene's test for homogeneity, Pearson's correlation coefficient (Pearson's r) will be the chosen method to quantify the correlation, as it is suitable for normally distributed data. However, should one or both datasets deviate from normality, the Spearman rank correlation coefficient (Spearman's \(\rho\)) will be utilised, as it is robust to non-normally distributed data \citep{Sereno_2023}.

This methodical approach ensures that the choice of correlation function aligns with the nature of the data, ultimately providing a robust and statistically sound analysis of the correlation between GDELT and TextBlob sentiment analysis in the context of Hong Kong news articles.

\paragraph{Normality Testing:}

\begin{figure}[ht]
    \centering
    \includegraphics[width=0.95\textwidth]{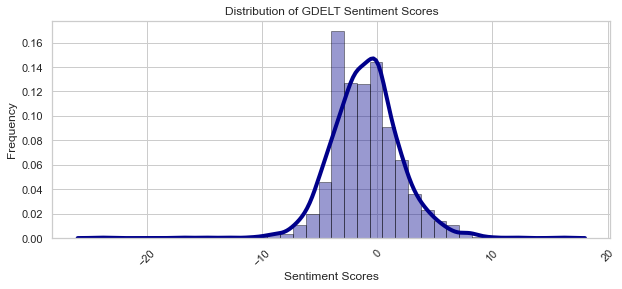}
    \caption{Distribution of GDELT Sentiment Scores}
    \label{fig:gdeltdist}
\end{figure}

\begin{figure}[ht]
    \centering
    \includegraphics[width=0.95\textwidth]{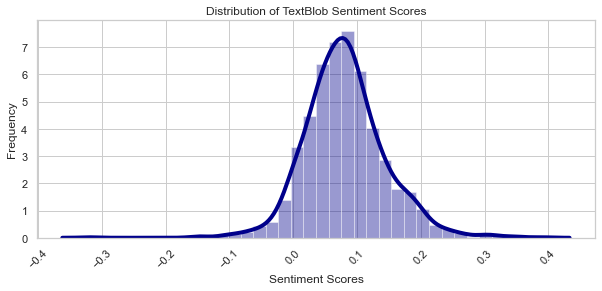}
    \caption{Distribution of TextBlob Sentiment Scores}
    \label{fig:tbdist}
\end{figure}

For both tone scores, the normality tests yielded p-values of less than 0.001, indicating a departure from normal distribution (the actual distributions for GDELT and TextBlob are depicted in Figure \ref{fig:gdeltdist} and Figure \ref{fig:tbdist} respectively). As a result, the data failed the Shapiro-Wilk test for normality. Given these deviations from normality, the Spearman rank correlation coefficient (Spearman's \(\rho\)) was be employed for the correlation analysis, as it is a non-parametric method robust to non-normally distributed data.

\paragraph{Spearman Rank Correlation Analysis:}
The Spearman correlation coefficient computed for the data is 0.427, indicating a moderate positive correlation between the variables under investigation. This value suggests that as one variable increases, the other tends to increase as well, though not in a perfectly linear fashion. The p-value associated with the Spearman correlation is highly significant, with a value of $<$0.001. This extremely low p-value provides strong evidence against the null hypothesis of correlation (which can thus be rejected), indicating that the observed correlation is statistically significant and not the result of random chance. 

While the correlation is not perfect, there is a significant positive correlation between the average tone scores obtained from GDELT and TextBlob sentiment analysis for Hong Kong news articles. As such, we continue to use the GDELT tonal scores for the rest of our analysis.





\subsection*{Hong Kong's Tone Trends (GDELT)}
As observed in Figure \ref{fig:tonegdelt}, there is a trend shift occurring in Hong Kong's average tone score in comparison to global averages. This shift saw Hong Kong moving from a position of significant tone score differences with the global trend to a period of reduced disparities from 2020 onwards. This shift underscores the need for hypothesis testing to comprehensively evaluate the statistical significance of these observations, beyond just visual analysis. The observed surge in average tone scores aligns temporally with the implementation of the NSL in Hong Kong.

\paragraph{Research Area:}
We seek to explore whether the implementation of the NSL in Hong Kong has coincided with the average tone scores of news articles (measured by GDELT) in a manner consistent with global trends (which would in effect mean, Hong Kong media's tone became relatively closer to the Chinese media's tone than previously) and major media outlets.

\paragraph{Null Hypothesis (H0):}
Following the implementation of the NSL in 2020, Hong Kong's average tone scores (GDELT) in news articles \textbf{diverged} from trends observed globally

\paragraph{Alternative Hypothesis (H1):}
Following the implementation of the NSL in 2020, Hong Kong's average tone scores (GDELT) in news articles \textbf{converged} towards trends consistent with global averages.

\paragraph{Interpretation and implications:} 
Recall that given the polarization of tones among countries, convergence towards the global (or cumulative) average essentially indicates a move away from being significantly negative, and reorienting to a more neutral or positive tone, in essence moving towards a more pro-China narrative.

\paragraph{Methodology:}
The primary focus was on tracking the average tone scores of news articles from the GDELT dataset, specifically computed on a monthly basis. To address potential variances between Hong Kong's sentiment trends and global counterparts, Welch's t-test \citep{welch_1947} was chosen for its suitability when dealing with groups exhibiting unequal variances \citep{Derrick_White_2016}.

The key analytical method employed was the rolling t-test, applied annually. This approach aimed to uncover specific timeframes when Hong Kong's sentiment trends deviated from the global norm, with a particular focus on determining if this divergence occurred post-2020. This methodology allowed for a systematic and statistically grounded exploration of the hypothesis, providing insights into whether the NSL had a discernible impact on the tone of news articles in Hong Kong and whether it led to a departure from global sentiment trends.\\

The rolling Welch's T-Test yielded the following results.
\begin{table}[h!]
    \centering
    \small
    \begin{tabu}{|l|[2pt]l|l|l|[2pt]l|l|l|l|}
    \hline
    \diagbox[width=\dimexpr \textwidth/4+4\tabcolsep\relax, height=1cm]{Country}{Year}
                              & 2017                      & 2018                      & 2019                      & 2020           & 2021                      & 2022           & 2023                      \\ \tabucline[2pt]{-}
    Global         & 0.449                     & \textbf{\textless{}0.001} & \textbf{0.020}                     & 0.692          & 0.518                     & 0.409          & 0.148                     \\ \hline
    China          & \textbf{\textless{}0.001}                     & \textbf{\textless{}0.001} & \textbf{\textless{}0.001} & \textbf{0.007} & \textbf{\textless{}0.001} & 0.064          & \textbf{\textless{}0.001} \\ \hline
    United States  & 0.128                     & \textbf{0.027}            & 0.143                     & 0.263          & 0.101                     & \textbf{0.015} & 0.157                     \\ \hline
    India          & \textbf{\textless{}0.001} & \textbf{0.002}            & 0.073                     & 0.326          & 0.067                     & \textbf{0.012} & \textbf{0.039}            \\ \hline
    United Kingdom & 0.795                     & 0.998                     & 0.741                     & 0.478          & \textbf{0.048}            & \textbf{0.003} & 0.077                     \\ \hline
    Malaysia       & \textbf{0.031}            & \textbf{0.017}            & 0.501                     & 0.101          & 0.099                     & 0.420          & \textbf{0.014}            \\ \hline
    Pakistan       & 0.413                     & \textbf{\textless{}0.001} & \textbf{0.027}            & 0.215          & 0.069                     & 0.828          & \textbf{0.004}            \\ \hline
    \end{tabu}
    \caption{Results of Welch's T-Test p-values by year}
    \label{tab:ttestresults}
\end{table}

\paragraph{Hong Kong vs cumulative Global average:}
The data from 2018 to 2019 shows significant difference in the average tone scores of news articles in Hong Kong when compared to the global average tone. This divergence in sentiment may be attributed to the build-up and events surrounding the Hong Kong protests during those years. However, the data suggests a notable shift in 2020 and subsequent years, with p-values indicating no significant difference between Hong Kong and global average tones. This transition may reflect the impact of the NSL and subsequent media control measures, effectively aligning the tone of Hong Kong's news articles more closely with cumulative average.

\paragraph{Hong Kong vs China:}
Notwithstanding that Hong Kong's tone changed and aligned closer to the cumulative average, the data from 2017 to 2023 shows that Hong Kong's tone did not however converge with China's tone. While 2022 had a close p-value of 0.064, indicating some similarity, the wider analysis suggests that Hong Kong maintained a distinct tone throughout these years. 

\paragraph{Hong Kong vs United Kingdom:}
The data from 2021 to 2023, with 2023 having a close p-value of 0.077, shows Hong Kong starting to show statistical differences from the United Kingdom. This indicates a potential shift in the trends of news article tones between Hong Kong and the United Kingdom, especially given that Hong Kong and the United Kingdom had similar average tones from 2017 to 2019 as seen in Figure \ref{fig:tonegdelt}.

\paragraph{Other inferences:}
Another notable shift is how Hong Kong transitioned from having statistical differences with countries that are closer to China, such as Malaysia and Pakistan, from 2017 to 2019. In contrast, from 2020 onwards, there are mostly no statistical differences. This indicates that Hong Kong is moving towards having similar trends with countries that are friendly to China, even though it has not (yet) converged with that of China itself.

\paragraph{Results:}
The results suggest that the null hypothesis does not hold for the examined period. As depicted in Table \ref{tab:ttestresults}, Hong Kong's average tone scores exhibited dynamic shifts. Initially, Hong Kong showed significant differences when compared to global trends in the years 2018 and 2019. However, following the implementation of the NSL in 2020, the tone scores converged towards trends consistent with global (cumulative) averages. Notably, this transition was marked by a lack of significant differences between Hong Kong and global tone scores from 2020 onwards. Furthermore, Hong Kong exhibited no significant differences with countries friendly to China such as Malaysia and Pakistan, indicating a potential shift in media sentiment aligning with countries close to China. 
All in all, we reject the null hypothesis and accept the alternative hypothesis that following the implementation of the NSL in 2020, Hong Kong's average tone scores (GDELT) in news articles \textbf{converged} towards trends consistent with global (cumulative) averages. We note that our study however only establishes the timing and change in trend, but does not prove causality, which is outside the scope and means of the current study.



\subsection*{Variance of Hong Kong News Articles}




Variance of tones for a given country acts as a proxy for the diversity in opinions being covered in its media. Nevertheless, examining Hong Kong's variance alone may not provide a comprehensive picture, e.g., if there are inherently many diverse aspects in a given time window which have individually been received more positively or negatively, this may also lead to larger variance. Thus comparing Hong Kong's variance with the global cumulative and of other countries' individual variances provides secondary insights into potential shifts in perspective. If and when Hong Kong's variance fluctuates (particularly if it is large) similarly as in other countries or with the global cumulative, it is more likely that it is a consequence of the inherent events around the BRI in that time frame, while changes in Hong Kong which are not in sync with the rest can be considered more indicative of changes in Hong Kong media's stance itself.



\begin{figure}[ht!]
    \centering
    \includegraphics[width=0.95\textwidth]{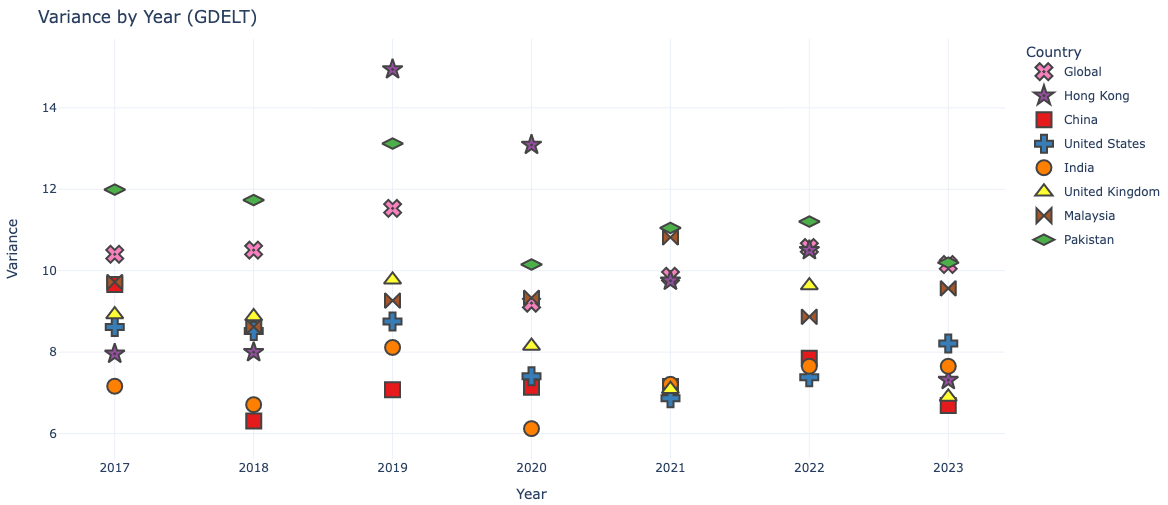}
    \caption{Variance per year by country (GDELT)}
    \label{fig:var}
\end{figure}

Figure \ref{fig:var} illustrates the variance in tone for the various countries across the years. 

\newpage

\paragraph{Hong Kong:} As observed in Figure \ref{fig:var}, the variance of Hong Kong's sentiment remained relatively stable and consistent in 2017 and 2018, indicating a certain degree of uniformity in the tone of news articles during those years. However, a substantial shift occurred in 2019, when the variance sharply increased. There was some increase in variance across all regions considered, nevertheless, Hong Kong experienced a particularly pronounced spike. Further text analysis (outside the scope of this work) is needed to determine the general global rise in variance and correlation with geopolitical events such as the US-China trade war around that time, while we do some further preliminary analysis specifically of the Hong Kong articles to get a sense of the source of this large spike, which we report below. The variance eventually decreased in the most recent years. This follows the enactment of National Security Law (2020) and coincides with the sharp fall in Hong Kong's press freedom \citep{Reuters} as seen in Figure \ref{fig:hkgraph}, however, our study is inadequate to establish any causality even while we remark on the correlation.  




\paragraph{Other inferences:} 

India and China have some of the lowest variance for most of the years, which suggests that the media in each of these countries have more coherent coverage of the BRI. For China, it is readily understandable since most of the news organizations are state run, while it appears that the Indian media, despite its variety in terms of sources (as previously observed in Figure \ref{fig:indist}), are consistently critical (since they have one of the most negative tone, as discussed previously in the context of Figure \ref{fig:tonegdelt}). In recent years, United Kingdom and United States also have very low variance - in combination with the observation from Figure \ref{fig:tonegdelt} that they both now have highly negative tones, a reasonable inference is that not withstanding the large diversity in their media and political landscapes, there is increasing coherence in BRI skepticism. This concurs well with the general trends in these countries, for instance, China skepticism is one of few aspects that Republicans and Democrats largely agree on \citep{wapo}.    

Pakistan on the contrary, while overall positively disposed towards the BRI, has consistently high variance. Further analysis of the actual text (outside the scope of this work) is necessary, some likely candidate causes include concerns of potential debt and transparency issues \citep{PToday}, environmental impact as well as skepticism of the native population \citep{baloch} of the restive Balochistan region in Pakistan which is adversely affected by CPEC. Irrespective of the causes, the high variance indicates a possibly vibrant and diverse tone in coverage, notwithstanding the smaller set of media entities and political circumstances of the country.

\subsection*{Text Analysis}
The dip in sentiment tone between 2018 and 2019 as seen in Figure \ref{tab:ttestresults} and the large variance observed in 2019 and 2020 from Figure \ref{fig:var} warrants investigation to explore possible connection to the Hong Kong protests. An additional focal point of this text analysis will be the examination of whether references to the National Security Law in these articles correlate with a lower sentiment score, clarifying the complex relationships that underlie our sentiment analysis results. 

\subsection*{Hong Kong Riots (2019)}
\paragraph{Analysis for 2018:} In 2018, the majority of articles that mentioned `riots' were primarily focused on the Xinjiang region in China. These articles likely discussed the unrest and social issues within Xinjiang. On the other hand, articles mentioning `protests' in the same year predominantly covered global protest movements in relation to the BRI's projects. The BRI's ambitious infrastructure projects sparked worldwide demonstrations and opposition, which were widely covered in the media.

In 2018, the 65 articles mentioning `protest' or `riot' had an average tone of -3.85, while the entire dataset of 602 articles had an average tone of -0.905. Removing these articles from the dataset and recalculating the average tone results in a significant improvement, with the average tone becoming -0.593.

Therefore, while it is challenging to fully attribute the sentiment divergence for the year 2018 to a single cause, it is evident that prominent media coverage of riots in Xinjiang and global protests related to the BRI played a significant role, yet, for 2018, there is inadequate evidence to link the negative sentiments to the domestic turmoils within Hong Kong itself.

\paragraph{Analysis for 2019:}
In 2019, a mere 13 articles mentioned the Hong Kong riots, and upon analysis, it was evident that they did not exert a significant influence on the average sentiment score. Consequently, it becomes challenging to attribute the shift in sentiment in 2019 to the media coverage of the Hong Kong riots in relation to the BRI. Notably, it can therefore be inferred that most articles covering the BRI primarily center on the initiative itself and remain relatively impervious to geopolitical events related to China, even major occurrences such as the Hong Kong riots.

\subsection*{National Security Law}
Filtering through the articles mentioning the NSL resulted in the following distribution of average tone and number of articles.

\begin{table}[h!]
    \centering
    \small
    \begin{tabu}{|l|[2pt]l|l|}
    \hline
    Year & Average Tone & Count \\ \tabucline[2pt]{-}
    2017 & -2.4         & 5     \\ \hline
    2018 & -5           & 2     \\ \hline
    2019 & NA           & 0     \\ \hline
    2020 & -4.25        & 4     \\ \hline
    2021 & 0.33         & 3     \\ \hline
    2022 & 2            & 2     \\ \hline
    2023 & 0.33         & 6     \\ \hline
    \end{tabu}
    \caption{NSL Articles: Average Tone and Count}
    \label{tab:nslvar}
\end{table}

While there are very few articles mentioning the NSL, this further demonstrates that articles discussing the BRI tend to be focused and are typically not directly influenced by geopolitical events within Hong Kong. However, it's essential to note that the absence of direct mentions does not necessarily imply that sentiments and behaviours are unaffected. The implementation of the NSL extends beyond mere news coverage; it significantly impacts how journalists and media outlets can address China, making it a nuanced aspect of media dynamics.

\section*{Concluding remarks}
\subsection*{Summary of Key Findings}

Our findings reveal a significant positive correlation between GDELT's sentiment analysis and more advanced sentiment analysis tools like TextBlob. This finding enhances the credibility of our study's experimentation and analysis, particularly since our study relies heavily on GDELT's average tone. 

While advanced sentiment analysis tools have the potential to provide superior and more comprehensive results, our methodology underscores how one can effectively analyse a topic without the need for extensive computational time or highly advanced NLP tools. This demonstrates the feasibility of our approach and its applicability in a wide range of research contexts using easily available resources such as GDELT.

Studying Hong Kong's average sentiment scores, we discern significant shift in the coverage of the BRI before and after the implementation of the NSL. Post-implementation, we observed a notable shift in average tone, aligning more closely with global cumulative sentiment, which in effect indicates a move in Hong Kong's tone closer to that of countries such as Pakistan and Malaysia which have friendly relations with China, even as it did not echo the tone of China itself.

We noticed an intermediate spike in variance for Hong Kong, which roughly coincided with a dip in the overall sentiment tone, indicating a period where more skeptical articles on the BRI were published around the same time when there was significant internal turmoils in Hong Kong, and yet those Hong Kong specific issues were generally not mentioned in the articles (see below remarks on text analysis) so it is non-conclusive whether the increased negative tone was implicitly motivated by them or not. However, subsequent to the National Security Law of 2020, in recent years one can observe that there are few articles on the BRI, they have a reasonably neutral tone, and with low variance. 

In the text analysis, a notable key finding was the absence of significant coverage on geopolitical issues in Hong Kong and the National Security Law within the BRI narratives in Hong Kong media outlets.
This observation indicates that, within the scope of their reporting on the BRI, Hong Kong media outlets generally refrained from incorporating or extensively discussing geopolitical matters specific to Hong Kong, particularly in the context of the National Security Law and the Hong Kong riots. This finding underscores a distinct separation between the coverage of the BRI and the local geopolitical issues in Hong Kong within the media landscape, suggesting that these topics are generally treated separately or addressed with limited overlap in Hong Kong media narratives.

It is crucial to emphasise that while these trends and findings are evident, we cannot directly infer that the implementation of the National Security Law caused this shift. We emphasize that correlation does not inherently imply causation, and our methodology does not facilitate the establishment of causation. Further investigation is essential to that end. 


\subsection*{Geopolitical data science \& future directions}
    


Although this study provided valuable insights, it is important to acknowledge its limitations. The analysis was conducted solely on English news articles, while the demographics of Hong Kong extend beyond this linguistic scope, and there are many other BRI stakeholders that we did not study as a consequence. Notably, Italy very recently (6 December on 2023) exited from the BRI. Studying the Italian media landscape over the period from before it joined till it exited the BRI, exploring the key themes over time, would make an interesting case study to drive with non-English content. Moreover, the application of advanced NLP tools and text analysis techniques can be considered to delve deeper into the factors underlying the observed shift in sentiment, for example, in an attempt to establish causation. 

Beyond the subject matter of immediate focus in this study, our work demonstrates how the methodology employed can be readily transferred to other topics, extending the pathway for data science applied to geopolitical study. In the process, it also provides a validation of GDELT's tone sentiments as a valuable tool for conducting such insightful research. While more advanced NLP tools are available, our study illustrates that GDELT's tone sentiment is a ready to use resource for a quick start, even as one might then complement it with advanced NLP techniques.

\bibliography{reference}

\newpage
\appendix
\section*{Appendix: Non-English articles from media in NATO countries}

\begin{table}[!ht]
    \centering
    \footnotesize
    \begin{tabu}{|l|[2pt]l|l|l|l|l|l|l|}
    \hline
    \diagbox[width=\dimexpr \textwidth/4+4\tabcolsep\relax, height=1cm]{ Country }{Year}
                   & 2017 & 2018 & 2019 & 2020 & 2021 & 2022 & 2023 \\ \tabucline[2pt]{-}
        Albania & 5 & 2 & 0 & 0 & 0 & 0 & 0 \\ \hline
        Belgium & 0 & 2 & 0 & 0 & 0 & 0 & 0 \\ \hline
        Bulgaria & 2 & 2 & 0 & 0 & 0 & 0 & 0 \\ \hline
        Canada & 39 & 17 & 11 & 14 & 5 & 1 & 0 \\ \hline
        Croatia & 1 & 0 & 0 & 0 & 0 & 0 & 0 \\ \hline
        Czech Republic & 5 & 4 & 3 & 0 & 0 & 3 & 0 \\ \hline
        Denmark & 0 & 0 & 0 & 0 & 0 & 0 & 0 \\ \hline
        Estonia & 0 & 0 & 0 & 0 & 0 & 0 & 0 \\ \hline
        Finland & 2 & 0 & 0 & 0 & 0 & 0 & 0 \\ \hline
        France & 168 & 8 & 3 & 5 & 4 & 1 & 0 \\ \hline
        Germany & 34 & 22 & 27 & 6 & 6 & 0 & 0 \\ \hline
        Greece & 12 & 2 & 5 & 1 & 1 & 0 & 0 \\ \hline
        Hungary & 4 & 1 & 0 & 0 & 0 & 0 & 0 \\ \hline
        Iceland & 0 & 3 & 0 & 0 & 0 & 0 & 0 \\ \hline
        Italy & 16 & 6 & 11 & 0 & 1 & 0 & 0 \\ \hline
        Latvia & 7 & 1 & 0 & 0 & 0 & 0 & 0 \\ \hline
        Lithuania & 0 & 0 & 0 & 0 & 1 & 0 & 0 \\ \hline
        Luxembourg & 4 & 0 & 0 & 0 & 0 & 0 & 0 \\ \hline
        Montenegro & 0 & 0 & 0 & 0 & 0 & 0 & 0 \\ \hline
        Netherlands & 1 & 4 & 9 & 1 & 0 & 0 & 0 \\ \hline
        North Macedonia & 0 & 1 & 0 & 0 & 0 & 0 & 0 \\ \hline
        Norway & 0 & 1 & 1 & 0 & 0 & 0 & 0 \\ \hline
        Poland & 4 & 1 & 0 & 0 & 0 & 0 & 0 \\ \hline
        Portugal & 2 & 3 & 1 & 1 & 0 & 0 & 0 \\ \hline
        Romania & 0 & 0 & 0 & 0 & 0 & 0 & 0 \\ \hline
        Slovakia & 1 & 0 & 0 & 0 & 0 & 0 & 0 \\ \hline
        Slovenia & 3 & 0 & 0 & 0 & 0 & 0 & 0 \\ \hline
        Spain & 3 & 10 & 4 & 2 & 1 & 0 & 0 \\ \hline
        Turkey & 42 & 8 & 15 & 0 & 2 & 0 & 0 \\ \hline
        United Kingdom & 4 & 2 & 0 & 1 & 0 & 0 & 0 \\ \hline
        United States & 11 & 1 & 2 & 1 & 0 & 0 & 0 \\ \hline
    \end{tabu}
    \caption{Number of non-English articles on the BRI identified from NATO member countries}
    \label{tab:natooutput}
\end{table}

\end{document}